\documentclass[12pt]{article}

\begin{document}
\begin{center}
{\Large \bfseries The mass of the $\tau$ neutrinos}

\vspace{.7cm}
\large{E.L. Koschmieder}
\medskip

\small{Center for Statistical Mechanics, The University of Texas at 
Austin\\Austin, TX 78712 USA\\e-mail: koschmieder@mail.utexas.edu}
\smallskip
\end{center}

\bigskip
\noindent
\small{We have shown previously that the mass of the muon neutrino can be 
determined from the energy released in the decay of the $\pi^\pm$ mesons, 
and that the mass of the electron neutrino can be determined from the 
energy released in the decay of the neutron. We will now show how the mass 
of the tau neutrino can be determined from the decay of the $D^\pm_s$ 
mesons.}

\normalsize

\section{Introduction}

   As we have shown with the standing wave model of the stable mesons and 
baryons [1] it follows from the decay of the $\pi^\pm$ mesons that 
the mass of the muon neutrinos $\nu_\mu$ and $\bar{\nu}_\mu$ must be 
m($\nu_\mu$) = m($\bar{\nu}_\mu$) = 47.5 
meV/$c^2$. We have also found in [1] that it follows from the decay of the 
neutron that the mass of the antielectron neutrino $\bar{\nu}_e$ should be 
m($\bar{\nu}_e$) = 0.55 meV/$c^2$. We have to correct this value whose 
calculation was based on the assumption that the neutron, whose mass is 
$\approx$ 2$K^\pm$, is the superposition of a $K^+$ and a $K^-$ meson. 
However, our investigation of the spin of the mesons and baryons [2] has 
shown that such a superposition does not produce spin s = 1/2, as is 
required. On the other hand, the superposition of two $K^0$ mesons has 
spin s = 1/2; actually the neutron must be the superposition of a $K^0$ 
and a $\bar{K}^0$ meson, because of conservation of strangeness in the 
strong interaction that created the neutron. According to the standing 
wave model the superposition of a $K^0$ and a $\bar{K}^0$ consists of 
neutrinos only and their oscillation energy. The neutrinos are arranged in 
a cubic lattice, each cell containing $\nu_\mu, \bar{\nu}_\mu, \nu_e, 
\bar{\nu}_e$ neutrinos. The total number of the neutrinos in the neutron 
is 4N, with N = 2.854$\cdot 10^9$, twice as many neutrinos as if the 
neutron were a superposition of $K^+$ and $K^-$. That means that, when the 
neutron decays via n $\rightarrow$ p + $e^-$  + $\bar{\nu}_e$, N 
antielectron neutrinos $\bar{\nu}_e$ share the difference $\Delta$ 
of the energy in the rest mass of the neutron and the energy in the rest 
mass of the proton $\Delta$ = 1.29332 MeV. After the rest mass of the 
electron, also emitted in 
the decay of the neutron, is subtracted from $\Delta$, and $\Delta$ 
$\mathrm{-}$ m(e)$c^2$ = 
0.782321 MeV is divided by N, not by N/2 as in [1], we find that the mass 
of the 
antielectron neutrino must be 

\begin{equation} \mathrm{m} (\bar{\nu}_e) = 0.275\,\mathrm{meV/c^2}\,. 
\end{equation}

    From the decay 
of the antineutron $\bar{n}$ $\rightarrow$ $\bar{p}$ + 
$e^+$  + $\nu_e$ follows in principle that the mass of the electron neutrino should be 
m($\nu_e$) = 0.275\,meV/$c^2$, or that m($\nu_e$) = m($\bar{\nu}_e$).

\section{The mass of \,$\nu_\tau$}

   We will now determine the mass of the $\tau$ neutrino m($\nu_\tau$) 
from the decay of the $D^\pm_{s}$ mesons in a way which is analogous to 
the 
way how we have determined the mass of $\nu_\mu$. The $D_{s}^\pm$ mesons 
decay via e.g.

\begin{equation} D^+_{s}\rightarrow \tau^+ + \nu_\tau \,\,(6.4\%)\,, \end 
{equation}
   where $\tau^+$ is the positive $\tau$ meson or lepton. The decay of 
$D^-_s$ has the conjugate particles on the right hand side of (2). The 6.4 
percentage of this mode of decay is, by a small margin, the most frequent 
of the leptonic modes of decay of $D^\pm_s$. The subsequent decay of 
$\tau^\pm$ is given by e.g.

\begin{equation} \tau^+ \rightarrow \pi^+ + \bar{\nu}_\tau\,\,(11.06\%)\,,
\end{equation}
 with the antitau neutrino $\bar{\nu}_\tau$. The percentage of this mode 
of decay of $\tau^\pm$ is likewise only one of the very many modes of 
decay of $\tau^\pm$.

   If it is true, as we have postulated in [1], that the particles consist 
of the particles into which they decay, then it follows from Eqs.(2,3) 
that the $D^\pm_s$ mesons consist of $\nu_\tau$ and $\bar{\nu}_\tau$ 
neutrinos, plus the $\nu_\mu, \bar{\nu}_\mu, \nu_e, \bar{\nu}_e$ neutrinos 
in the $\pi^\pm$ mesons in Eq.(3). The cells of the lattice of the 
$D^\pm_s$ 
mesons contain 6 types of neutrinos, not 4 types as in the $\pi^\pm$ 
mesons. The cubic lattice used in the standing wave model can, however, be 
retained if we consider, instead of a simple cubic lattice as the NaCl 
lattice, a body-centered cubic lattice in which a particle different from 
the particles in the corners of the simple cubic cell sits at the center 
of each cell of the lattice. This seems to accomodate only 5 neutrino 
types, whereas we have found that there must be 6 neutrino types in the 
cells of the $D^\pm_s$ mesons. However, because of conservation of lepton 
numbers during the creation of the $D^\pm_s$ mesons it is necessary that a 
number of antitau neutrinos equal to the number of tau neutrinos is 
present in the lattice. The antitau neutrinos can easily be accomodated in 
a lattice of body-centered cells in which the center particles are 
alternately tau neutrinos and antitau neutrinos.

   As explained in [1] there must be N = 2.854$\cdot10^9$ neutrinos in the 
$\pi^\pm$ lattice, and consequently there are N/4 simple cubic cells in 
$\pi^\pm$. The number of cells in the $\pi^\pm$ lattice and the lattice of 
the neutron seem to be the same, because the radii of the $\pi^\pm$ mesons 
and the proton are, within the accuracy of the measurements, the same, and 
we assume that the size of the proton and neutron are the same. It appears 
that the superposition of a $K^0$ and a $\bar{K}^0$ meson creating a 
neutron does not change the size of the lattice of  these particles, or 
the number of their cubic cells. We will therefore assume that the 
superposition of a proton, an antineutron and a $\pi^0$ meson creating the 
$D^\pm_s$ mesons, with m($D^\pm_s$) = 0.978$\cdot$(\,m(p) + m($\bar{n}$) + 
m($\pi^0$)\,), does not change the number of the cells in the lattice 
either. In other words we assume that the size of the proton or neutron is 
the same as the size of the $D^\pm_s$ mesons. If there are N/4 
body-centered cells in the $D^\pm_s$  meson then there must be N/8 tau 
neutrinos and antitau 
neutrinos each in the lattice of $D^\pm_s$. The energy $\Delta$ released 
in the decay $D^+_s$  $\rightarrow$  $\tau^+$ + $\nu_\tau$ (Eq.(2)) is 
given by 

\begin{equation} \Delta = \mathrm{m}(D^\pm_s)c^2\, \mathrm{-}\,
\mathrm{m}\,(\tau^\pm)c^2 = 191.51\, \mathrm{MeV}\,.\end{equation}
If this energy originates from the rest mass of all $\nu_\tau$ neutrinos, 
respectively from all $\bar{\nu}_\tau$ neutrinos, in the decay of 
$D^\pm_s$ 
then it follows, with the number of $\nu_\tau$ or $\bar{\nu}_\tau$ 
neutrinos being N/8, that 

\begin{equation} \mathrm{m}(\nu_\tau) = \mathrm{m}(\bar{\nu}_\tau) = 
536.8\, \mathrm{meV/c^2} \approx 0.54\, 
\mathrm{eV/c^2}\,.\end{equation}
 
   Since in the decay of $D^+_s$ only a $\emph{single}$ tau neutrino is 
emitted one must wonder why the energy $\Delta$ released in the decay 
should be equal to the sum of the rest masses of $\emph{all}$ tau 
neutrinos 
in $D^+_s$. The first indication that $\Delta$ is not the energy carried 
by a single tau neutrino in the $D^+_s$ meson comes from the magnitude of 
$\Delta$ which amounts to practically 10\% of the energy in the rest mass 
of $D^+_s$. This is incompatible with the basic tenet that there must be, according 
to Fourier analysis, a continuum of frequencies in a body created in a 
high energy collision of $10^{-23}$ sec duration, which does not make it 
possible that a single neutrino out of $10^9$ neutrinos has a rest mass 
plus an oscillation 
energy amounting to 10\% of the rest mass of $D^\pm_s$. That the source of 
$\Delta$ is 
the rest masses of all $\nu_\tau$, respectively $\bar{\nu}_\tau$,
neutrinos can be inferred from the disappearance of one of the two 
neutrino types in 
the secondary decays following the primary decay of $D^\pm_s$. To be 
specific, the $\nu_\tau$ in the decay of $D^+_s$ (Eq.(2)) does not appear 
neither in the 
decay of $\tau^+$ (Eq.(3)) nor in the subsequent decay of $\pi^+$  which 
is in also (3), nor in 
the decay of the $\mu^+$ meson which follows. With the primary decay of 
$D^+_s$, 
(Eq.(2)), the tau neutrinos seem to have been eliminated, which will 
certainly be the case when the energy of the rest masses of all 
$\nu_\tau$ has been consumed by $\Delta$. This process is analogous to the 
disappearance of one type of muon neutrino after the decay of the 
$\pi^\pm$ mesons, as discussed in [1], where we have shown that the 
oscillation energy of all neutrinos in $\pi^\pm$ is conserved in the 
decay, where therefore the energy in the emitted neutrino can come only 
from the sum of the rest masses of all neutrinos of the type of neutrino 
emitted in the decay.

   Finally we want to show that the body-centered lattice of the $D^\pm_s$ 
meson leads also to the correct spin s($D^\pm_s$) = 0. We have shown in 
[2] and [3], e.g. in context with the spin of the $\pi^\pm$ mesons 
s($\pi^\pm$) = 0, that the spin of the electric charge carried by the 
$\pi^\pm$ mesons is canceled by the sum of the spin vectors of all 
neutrinos in the lattice. Of all the spin vectors of the O($10^9$) 
neutrinos in the cubic lattice of the $\pi^\pm$ mesons only the spin 
vector of the central neutrino remains, which then cancels the spin vector 
of the electric charge. The same applies for the body-centered lattice of 
the $D^\pm_s$ mesons. Of the spin vectors of all neutrinos in the 
$D^\pm_s$ lattice only the spin vector of the central neutrino, in this 
case of a $\nu_\tau$ or $\bar{\nu}_\tau$ neutrino, remains which cancels 
the spin vector of the electric charge of $D^\pm_s$. Consequently 
s($D^\pm_s$) = 0, as it must be.

\section{Conclusions}

   Making use of the decay of the $D^\pm_s$ mesons $D^\pm_s$ $\rightarrow$ 
$\tau^\pm$ + $\nu_\tau(\bar{\nu}_\tau)$  we can show that 
the mass of the tau neutrino must be m($\nu_\tau$) $\approx$  
0.54\,eV/$c^2$. We also find that m($\nu_\tau$) = m($\bar{\nu}_\tau$).
 
\bigskip
\bigskip
\noindent
\textbf{REFERENCES}

\smallskip
[1]\,E.L.Koschmieder, arXiv:physics/0211100 (2002),    \\
\indent 
\quad\,Chaos, Solitons and Fractals {\bfseries18},1129 (2003).

\smallskip
[2]\,E.L.Koschmieder, arXiv:physics/0301060 (2003), Hadr.J. (to appear).

\smallskip
[3]\,E.L.Koschmieder, arXiv:physics/0308069 (2003).

\end{document}